\documentclass[prd,twocolumn,aps,amsmath,amssymb,nofootinbib,preprintnumbers]{revtex4}

\usepackage{graphicx}
\usepackage{dcolumn}
\usepackage{bm}
\usepackage{amsmath}
\usepackage{amsfonts}

\def\drawbox#1#2{\hrule height#2pt
        \hbox{\vrule width#2pt height#1pt \kern#1pt
              \vrule width#2pt}
              \hrule height#2pt}

\def\Asym#1#2{\vcenter{\vbox{\drawbox{#1}{#2}
              \kern-#2pt       
              \drawbox{#1}{#2}}}}

\newcommand{\beq}{\begin{eqnarray}}
\newcommand{\eeq}{\end{eqnarray}}

\usepackage{verbatim}

\newcommand{\p}{\partial}

\newcommand{\vp}{\varphi}

\newcommand{\dsigma}{\delta^{\textrm{\tiny (1)}} \sigma}
\newcommand{\nn}{\nonumber \\}

\newcommand{\Hcal}{{\cal H}}
\newcommand{\Jcal}{{\cal J}}
\newcommand{\ixo}{\textrm{\tiny (1)}}
\newcommand{\ixt}{\textrm{\tiny (2)}}

\begin{document}

\title{Non-Gaussianity from Preheating}
\author{Kari Enqvist$^{1,3}$}
\author{Asko Jokinen$^{2}$}
\author{Anupam Mazumdar$^{2}$}
\author{Tuomas Multam\"aki$^{2}$}
\author{Antti V\"aihk\"onen$^{1,3}$}
\affiliation{$^{1}$Helsinki Institute of Physics, P.O. Box 64, FIN-00014 University of
Helsinki, Finland\\
$^{2}$NORDITA, Blegdamsvej-17, Copenhagen-2100, Denmark.\\
$^{3}$Department of Physical Sciences, University of Helsinki,
P.O. Box 64, FIN-00014 University of Helsinki, Finland}

\begin{abstract}
We consider a two-field model for inflation where the second order
metric perturbations can be amplified by a parametric resonance during
preheating. We demonstrate that there can arise a considerable
enhancement of non-Gaussianity sourced by the local terms generated
through the coupled perturbations. We argue that the non-Gaussianity
parameter could be as large as $f_{NL}\approx 50$. Our results
may provide a useful test of preheating in future CMB experiments.
\end{abstract}
\preprint{HIP-2004-62/TH}
\preprint{NORDITA-2004-91}
\maketitle

{\it Introduction:} Preheating, first realized in~\cite{Robert} and
worked out in detail in~\cite{Many1,Linde,Dan}, may play an important
role in understanding hot thermal Universe after the end of inflation.
Preheating can occur during the coherent oscillation of the
homogeneous inflaton condensate if there exists a temporary vacuum
instability due to which a resonant production of particles takes
place.  During this phase it is also possible to amplify the
gravitational fluctuations or the metric fluctuation to the super
Hubble scale~\cite{Many2,Liddle-Wands} (for a general first order
linear perturbation theory, see~{\cite{MFB}).  So far the analysis has
been limited to the first order metric and matter perturbations,
although there has also been some attempts to understand the higher
order effects~\cite{Easther}. In some case the inflaton could
also fragment into non-topological solitons, but then the
inhomogeneities remain below the subhorizon scale~\cite{Kas1}.

Since CMB experiments such as WMAP~\cite{WMAP} have now reached a
precision whereby cosmological models can be put to test,
preheating may also move from the theoretical playground to
observational scrutiny. In the present paper we propose that
non-Gaussianities in the CMB fluctuations could provide a useful
tool for studying preheating.

The simplest single field inflationary models produce scale invariant
Gaussian fluctuations, assuming that the initial state is the standard
vacuum. However, non-Gaussianities are expected to be generated at
some level, either because of inflaton
self-couplings~\cite{Maldacena}, non-standard initial vacuum
state~\cite{Gangui}, or because of a host of other reasons often
involving models beyond the simple single field
inflation~\cite{Many3,Uzan,Antti} (for a review, see
\cite{ngrev}). For example, in hybrid-type scenarios with two fields
the fluctuations of the "waterfall" field may give rise to large
non-Gaussianities during inflation~\cite{Antti}.

Major non-Gaussianities may also be generated after
inflation~\cite{Riotto}.  Here we focus on non-Gaussianities during
the preheating stage, although our approach is quite generic
and applicable to any scalar condensate responsible for reheating the
Universe, such as a supersymmetric flat direction~\cite{Enqvist},
for which we may employ the formalism developed in~\cite{Antti}. We
will consider a simple toy model of two fields, making some
well-motivated physical assumptions regarding the motion of the
coupled fields. Our starting point is an oscillating inflaton coupled
to another field, which for simplicity is assumed to have a vanishing
vacuum expectation value. Therefore, classically the second field is
sitting at the bottom of its potential. However, its fluctuation about
the background solution is non-vanishing, giving rise to matter
perturbations which naturally are higher order, mainly chi-squared
fluctuations~\cite{Liddle}, which may source the metric perturbations
and lead to non-Gaussianities. We illustrate that although the first
order metric perturbations do not grow, for a long wavelength
super-Hubble horizon modes the second order metric fluctuations can
grow exponentially, highlighting the potential for a large
non-Gaussian signature arising from preheating.

For simplicity, for obtaining the estimates we will neglect the
expansion of the universe. This may seem rather restrictive, but since
the homogeneous condensate oscillates coherently with a frequency
larger than the Hubble expansion rate during the initial stages of
preheating, the only relevant time scale is the mass of the coherently
oscillating field. Of course, for detailed numerical values, the
expansion as well as backreaction effects will be important.


{\it Basic Equations:} Let us consider a two-field model with the potential
\begin{equation}
\label{maineq}
  V = \frac{1}{2} m_{\varphi}^2 \varphi^2 + \frac{1}{2} g^2 \varphi^2
  \sigma^2\,,
\end{equation}
where $\varphi$ is the homogeneous scalar condensate coherently
oscillating with a mass $m_{\varphi}$. We assume $\varphi$ to be our
inflaton. The $\sigma$ field will be created resonantly by virtue of
the coupling term.

For simplicity and for the sake of clarity we assume that the VEV of
$\sigma$ vanishes, $\langle\sigma\rangle=0$, which makes it possible
to obtain analytic approximations from the second order perturbation
equations, as we show below.  Such a situation occurs if the $\sigma $
field is driven to the minimum of its potential during inflation.

The metric in our case is given by
\begin{eqnarray}
\label{metric}
g_{00} &=& - a(\eta)^2 \left( 1 + 2\phi^{(1)} + \phi^{(2)} \right)\,, \nn
g_{0i} &=& 0 \,,\nn
g_{ij} &=& a(\eta)^2 \left( 1 - 2\psi^{(1)} - \psi^{(2)} \right) \delta_{ij}\,,
\end{eqnarray}
where we are using the generalized longitudinal gauge and neglect the
vector and tensor perturbations. Here $\eta$ is the conformal time and
$a(\eta)$ is the scale factor. We divide the fields into background and
perturbations,
\begin{eqnarray}
\label{split}
\varphi &=& \varphi_0(\eta) + \delta^{(1)}\varphi(\eta,{\bf x}) + \frac{1}{2}
\delta^{(2)}\varphi(\eta,{\bf x})\,, \nn
\sigma &=& \delta^{(1)}\sigma(\eta,{\bf x}) + \frac{1}{2}
\delta^{(2)}\sigma(\eta,{\bf x}) \,,
\end{eqnarray}
where the background value for $\sigma$ is assumed to vanish.

The background equations of motion are found to be
\begin{eqnarray}
\label{eqom}
3 {\cal H}^2 &= &\frac{1}{2M_{p}^2} \varphi_0'^{\,2} + \frac{M_{p}^2}{2}a^2
m_{\varphi}^2 \varphi_0^2 \,,\\
0&=& \varphi_0'' + 2 {\cal H} \varphi_0' + a^2 m_{\varphi}^2 \varphi_0\,,
\end{eqnarray}
while the $\sigma_0$-equation is trivial, and $\cal H$ denotes the
Hubble expansion rate in conformal time. We denoted the reduced
Planck mass as $M_{p}\sim 2.4\times 10^{18}$~GeV.

The relevant 1st order perturbation equations can be written in the
form~\cite{Antti}
\begin{eqnarray}
\phi^{(1)\,''} - \partial_i \partial^i \phi^{(1)} + 2 \left( {\cal H} -
  \frac{\varphi_0''}{\varphi_0'} \right) \phi^{(1)\,'} & & \nonumber \\ + 2
\left( {\cal H}' - \frac{\varphi_0''}{\varphi_0'} {\cal H} \right) \phi^{(1)}
&=& 0 \,,\label{bardeeneq} \\
\delta^{(1)}\sigma'' + 2 {\cal H} \delta^{(1)}\sigma' - \partial_i \partial^i
\delta^{(1)}\sigma & & \nonumber \\
+ g^2 \varphi_0^2 \, \delta^{(1)}\sigma &=& 0\,.
\label{chieq}
\end{eqnarray}
Note that there are no metric perturbations in Eq.~(\ref{chieq}). This
is due to assuming a vanishing VEV for $\sigma$. If it were
non-vanishing, we would not be able to decouple the metric
fluctuations from the perturbations in $\sigma$ field.  Now the
$\sigma$ part can be solved separately and for the rest the usual one
field results apply.

\vskip15pt
{\it Parametric Resonance:}
Let us now turn to the parametric resonance for $\dsigma_{\boldsymbol
k}$. Throughout this section we neglect the expansion of the Universe,
which results in $a=1$ and $\eta=t$, {\it i.e.} conformal and cosmic
time are equal, since the coherent oscillations begin only when
$m_{\varphi}\gg {\cal H}$. Hence the excitation of
$\sigma_{\boldsymbol k}$ will follow the equation of motion valid for
a {\it narrow parametric resonance} regime~\cite{Many2,Linde},
\begin{equation}
  \frac{d^2\dsigma_k}{dz^2} + (A_k - 2 q\, \mbox{cos}2z) \dsigma_k = 0\,,
\end{equation}
where $A_k=k^2/m_{\varphi}^2+2q$, $q = g^2 \Phi^2/4m_{\varphi}^2$, and
$z=m_{\varphi} t$. The amplitude of the oscillations is denoted by
$\Phi$.  In a narrow resonance regime $q<1$. In this regime the
perturbation grows exponentially as $\dsigma_k \propto \mbox{exp}
(\mu_k m_{\varphi} t)$, where
\begin{equation}
  \mu_k = \sqrt{(q/2)^2 - (2k/m_{\varphi} - 1)^2}~\,.
\end{equation}
Therefore, we have a resonance for the modes $k$ when $k_- < k < k_+$ with
$k_\pm  = (m_{\varphi}/2)(1 \pm q/2)$.

For our purposes it is sufficient to estimate the solution to be
independent of $k$ in the resonance band. Therefore, we estimate
$\dsigma_{\textrm{\tiny eff}} \equiv A\, \mbox{exp}
(\mu_{\textrm{\tiny eff}}m_{\varphi}t)$, where $\mu_{\textrm{\tiny
eff}} = \mu_{\textrm{\tiny max}}/2 = q/4$ in the resonance band and
$\dsigma_{\textrm{\tiny eff}} = 0$ otherwise; here $A$ is an amplitude
after the end of inflation.

The amplitude depends on the effective mass of $\sigma$ field. If
$g\varphi_0\leq m_{\varphi}$, then the initial perturbations is given
by $A\sim \delta^{(1)}\sigma_{k}\sim(H/\sqrt{2k^3})(k/aH)^{3/2-\nu}$
where $\nu=\sqrt{9/4-g^2\varphi_0^2/H^2}$. For a narrow resonance
regime, $g\leq H/\varphi_0\sim H/M_{p}\ll 1$, the spectrum follows
that of the inflaton, where we assumed at the end of inflation
$\varphi_0\sim M_{p}$.  For a broad resonance regime, $q>1$, and
$g\varphi_0\gg m_{\varphi}$, there the amplitude of the perturbations
for $\sigma$ field will be suppressed compared to that of the
inflaton, see~\cite{Liddle-Wands}.

\vskip10pt
{\it Perturbations:}
The first order metric perturbation at the end of inflation for a
single field is given by~\cite{MFB},
\begin{equation}
\phi^{(1)} = \frac{H}{\dot\varphi_0}\, \delta^{(1)}\varphi
\approx - \frac{\varphi_0}{2 M_{p}^2}\delta^{(1)}\varphi\,,
\label{phi1}
\end{equation}
where the quantities on the right-hand side are calculated at the
horizon crossing. We obtained the final result by using slow roll
equations of motion.  Since $g\ll 1$, then during inflation, we can
effectively treat $\delta^{(1)} \sigma \sim \delta^{(1)} \varphi$.
This determines the initial conditions before preheating starts.

In the second order we are only interested in the gravitational
perturbation, whose equation can be written in an expanding background
as~\cite{Antti}
\begin{align}
  & {\phi^{\ixt}}'' + 2 \left( \Hcal - \frac{\vp''_0}{\vp'_0} \right)
  {\phi^{\ixt}}' + 2 \left( \Hcal' - \frac{\vp''_0}{\vp'_0} \Hcal \right)
  \phi^{\ixt} \nonumber\\
  & - \p_i \p^i \phi^{\ixt} =
  \Jcal_{\sigma, \textrm{local}} + \Jcal_{\sigma, \textrm{non-local}}  +
  \Jcal_{\textrm{rest}}~\,, \label{eq:phi2}
\end{align}
where the source terms $\Jcal$ are quadratic combinations of first order
perturbations; in particular,
\begin{equation}
\label{local0}
  \Jcal_{\sigma, \textrm{local}} = -\frac{2}{M_{p}^2} 
  (\dsigma')^2 + \frac{a^2}{M_{p}^2}
  \frac{\delta^2 V}{\delta \sigma^2} (\dsigma)^2~\,.
\end{equation}
$\Jcal_{\sigma, \textrm{non-local}}$ involves an inverse spatial
Laplacian, thus rendering it non-local, while $\Jcal_{\textrm{rest}}$
consists of metric and $\varphi$ perturbations.  Note that the left
hand side of the above equation is identical to the first order
equation Eq.~(\ref{bardeeneq}).

Fourier transforming $\Jcal_{\sigma, \textrm{local}} \to \Jcal_k$
(with our convention $f(\boldsymbol x) = (1/2\pi)^3 \int d^3
\boldsymbol k \, e^{\boldsymbol k \cdot \boldsymbol x} f(\boldsymbol
k)$) we end up with the convolutions
\begin{align}
  & \Jcal_k = - \frac{2}{M_{p}^2(2\pi)^3} \int d^3 \boldsymbol k' \,
  \dsigma'_{\boldsymbol
    k'} \dsigma'_{\boldsymbol k - \boldsymbol k'} \nonumber\\
  & + \frac{1}{M_{p}^2}\frac{\delta^2 V}{\delta \sigma^2} 
  \frac{1}{(2\pi)^3} \int
  d^3 \boldsymbol k' \, \dsigma_{\boldsymbol k'} \dsigma_{\boldsymbol k -
    \boldsymbol k'}~\,.
\end{align}
The object is then to compute the convolutions. We are interested in
their contributions at large scales and, to that end, we take the
limit $k \to 0$, we also assumed $a=1$.  Since the mode function
$\dsigma_{\textrm{\tiny eff}}$ only depends on the magnitude of the
vector $\boldsymbol k$, the angular integration can be carried out
trivially. The time derivative only produces a constant factor.  Thus,
we obtain
\begin{align}
\label{align}
  \Jcal_k &= - 2 \frac{1}{M_{p}^2}\frac{4\pi}{(2\pi)^3} \int d k' {k'}^2
  (\dsigma'_{k'})^2 \nonumber\\
  & + \frac{a^2}{M_{p}^2} \frac{\delta^2 V}{\delta \sigma^2}
  \frac{4\pi}{(2\pi)^3} \int d k' {k'}^2 (\dsigma_{k'})^2 \\
  &= \Big[ -\frac{2\mu_{eff}^2 m_{\varphi}^2}{M_{p}^2}  + \frac{1}{M_{p}^2} 
  \frac{\delta^2 V}{\delta \sigma^2} \Big] \dsigma_{\textrm{\tiny eff}}^2
  \frac{4\pi}{(2\pi)^3} \int_{k_-}^{k_+} d k' \, {k'}^2~\,, \nonumber
\end{align}
where in the last step we have assumed that the $k$-dependence of the
amplitude $A$ can be ignored. If we are working in a narrow
resonance regime with $q<1$, the integral can be written as
\begin{equation}
\label{int}
  \int_{k_-}^{k_+} d k' \, {k'}^2 = \frac{1}{3} \left( \frac{m_{\varphi}}{2}
\right)^3
  \left[ 3q + 2 \left( \frac{q}{2} \right)^3
  \right] \simeq q \left( \frac{m_{\varphi}}{2} \right)^3~\,.
\end{equation}

We can now write the source term as
\begin{eqnarray}
  \Jcal_k &=& \frac{4\pi}{(2\pi)^3} q \left( \frac{m_{\varphi}}{2}
\right)^3 \left[ -\frac{q^2m_{\varphi}^2}{8M_{p}^2} + \frac{1}{M_{p}^2} 
\frac{\delta^2 V}{\delta \sigma^2} \right] A^2
e^{{qm_{\varphi}t}/{2}}
 \nonumber\\ &=& \frac{2m_{\varphi}^2 q}{M_{p}^2}
 \left[1-\frac{q}{16}\right]\times B
e^{{qm_{\varphi}t}/{2}}~\,. \label{C}
\end{eqnarray}
where $B=(q/8\pi^2)(m_{\varphi}/2)^3 A^2$.  It is worth noting that
the source $\Jcal_k$, which we study at the large scale limit $k \sim
0$, is actually generated by first order, local perturbations on much
smaller scales ($k_- < k < k_+$).

\vskip10pt
{\it Non-Gaussianity:}
Consider Eq.~(\ref{eq:phi2}) in $k$-space. The homogeneous part is the
same as in the first order. Therefore we know that the homogeneous
solutions are well-behaved. Barring accidental cancellations, we may
assume that the local terms we have considered are representative of
the exponential behavior of the source, see Eq.~(\ref{local0}). There
is also a non-exponential part which naturally becomes quickly
insignificant. In order to estimate the behavior of $\phi^{\ixt}$ at
large scales ($k \sim 0$) we neglect the expansion of the Universe and
drop the terms with $\Hcal$. The approximated metric perturbation then
reads
\begin{eqnarray}\label{metricappro}
  {\phi^{\ixt}_k}'' - 2 \frac{\vp_0''}{\vp_0'} {\phi^{\ixt}_k}' = 
\frac{2m_{\varphi}^2 q}{M_{p}^2}\left[1-\frac{q}{16}\right]\times B
  e^{{qm_{\varphi}t}/{2}} \nonumber \\
+ [\mbox{non-exponential source}]~\,,
\end{eqnarray}
Assuming that on average after some oscillations, the fraction
$\vp_0'' / \vp_0'$ can be approximated by the frequency of the
coherent oscillations $\sim m_{\varphi}$, we readily obtain an
exponential behavior for the solution of Eq.~(\ref{metricappro}). We
may thus write
\begin{equation}
\label{approx}
  \phi^{\ixt}_k \approx - \frac{2m_{\varphi}^2 
(1-q/16)\cdot B}{M_{p}^2 m_{\varphi}^2 (1- q/4)}
 e^{{qm_{\varphi}t}/{2}}~\,.
\end{equation}

Let us use the following definition for the constant non-linearity
parameter $f_{NL}^{\phi}$: $\phi = \phi^{\ixo} + f_{NL}^{\phi}
(\phi^{\ixo})^2$. However, the definition is given in $x$-space and we
have performed our calculations in $k$-space. In principle we could
transform $\phi^{(2)}_k$ back to $x$-space, but then we would need to
know it for all $k$ and we have evaluated only the superhorizon mode
$k=0$. Instead, we can carry the definition of $f_{NL}^{\phi}$ over to
$k$-space by $\phi^{(2)}_{\bf k} = f_{NL}^{\phi} \frac{1}{(2\pi)^3}
\int d^3 {\bf k'} \phi^{(1)}_{\bf k'} \phi^{(1)}_{{\bf k}-{\bf k'}}$
where we have treated $f_{NL}^{\phi}$ as a constant.

In the present scenario, where the 1st order perturbations are
equivalent to that of a single-field case, we have $\delta^{(1)}\sigma
\sim \delta^{(1)}\varphi$ right after inflation. Since, during
inflation and preheating $\phi^{\ixo}$ stays roughly constant, we
immediately obtain an order of magnitude estimate from
Eq.~(\ref{phi1})
\begin{eqnarray}
\label{fnl}
f_{NL}^{\phi} \sim \frac{\phi^{(2)}_k}{(\phi^{(1)}*\phi^{(1)})_k}
\approx -8\frac{1-q/16}{1-q/4} \left(\frac{M_{p}}{\varphi_0}\right)^2
e^{{Nq}/{2}}~\,,
\end{eqnarray}
where we have written $N=t\omega$, where $N$ is the number of
oscillations during preheating and $\omega$ is the frequency of the
oscillations. On the average the frequency of the oscillations is
given by $\omega\sim m_{\varphi}$.  Inflation ends when
$\varphi_{0}\sim M_{p}$, therefore the coefficient infront of the
exponential is order one. The factor $B$ in the coefficient of
Eq.~(\ref{approx}), whose origin lies in the source terms, see
Eqs.~(\ref{align},\ref{int},\ref{C}) cancels out completely from the
contribution coming from $(\phi^{(1)}\ast\phi^{(1)})_{\bf k}$, because
of the initial evolution for $\delta^{(1)}\varphi$ and
$\delta^{(1)}\sigma$ are the same. The above expression should be
compared with the observationally constrained one:
$f_{NL}=-f^{\phi}_{NL}+(11/6)$.

The amplitude of the oscillations remains constant in our case.
Therefore $N$ is an indicative number, which is valid until the
backreaction kicks in and shuts off the parametric resonance.
Obviously $N$ depends on a particular potential and also on the
expansion of the Universe. The parametric resonance would shift with
the expansion as $q\sim \Phi^2~t^{-2}$ and thus becomes
narrower. However, for a simple single field chaotic type inflation
model with $V\sim (m^2_{\varphi}/2)\varphi^2$ one would typically have
many oscillations within one Hubble time: $\omega H^{-1}\sim
M_{p}/\Phi \gg 1$. It seems therefore that backreaction would be more
decisive as far as the magnitude of the non-Gaussian amplitude
$f_{NL}^{\phi}$ is concerned. This requires more study, but let us
point out that in chaotic inflation backreaction becomes important
after 10-30 oscillations~\cite{Linde}. Hence $N={\cal O}(10)$ might be
a reasonable number, and should we for illustrative purposes choose
$q=0.8$, we would obtain $f_{NL}\approx e^{4}\approx 55$. This should
certainly be at an observable level for the Planck Surveyor Mission.

The expansion of the Universe changes the situation in two ways.
First, the parametric resonance can be broad with $q>1$ a time
dependent quantity~\cite{Linde}. Second, because of the the expansion
the momenta and the oscillation amplitude redshift. However the
amplitude of the first order metric perturbation still undergoes
resonant amplification as the momentum modes drift through the broad
resonance regime~\cite{Many2}. This ensures that the second order
metric perturbations also grow exponentially, but one has to
ensure that the amplitude of the initial perturbations for $\sigma$
does not damp away during inflation. A detailed study would
require numerical simulation, but nevertheless we may conclude that
our result hints at the possibility of exciting the second order
metric perturbations during the first few oscillations of the
inflaton, hence linking preheating with possibly observable
non-Gaussianities.



It is our pleasure to thank Andrew Liddle for discussions. K.E.~is
supported partly by the Academy of Finland grant no.~75065. A.V.~is
supported by the Magnus Ehrnrooth Foundation. A.V.~thanks NORDITA and
NBI for their kind hospitality during the course of this work,


\end{document}